\documentclass[12pt]{article}
\usepackage{amsxtra,amssymb,amsthm,amsmath,latexsym}

\theoremstyle{plain}

\def\nd{\noindent}

\def\R{{\mathbb R}}

\def\oH{\buildrel\circ\over H}
\def\oH1{\buildrel\circ\over H\kern-.02in{}^1}
\def\qed{{\hfill $\Box$}}
\def\l{\ell}

\begin{document}
J.Phys A, 35, (2002), L357-361.
%\begin{titlepage}

\title{ Modified Rayleigh Conjecture and Applications
   \thanks{key words: Rayleigh hypothesis, scattering by obstacles, 
inverse scattering
    }
}

\author{
A.G. Ramm\\
LMA/CNRS, Marseille 14302, cedex 20, France\\
and  Mathematics Department, 
Kansas State University, \\
 Manhattan, KS 66506-2602, USA\\
ramm@math.ksu.edu\\
}

\date{}

\maketitle\thispagestyle{empty}

\begin{abstract}
Modified Rayleigh conjecture (MRC) in scattering theory is proposed and
justified. MRC allows one to develop numerical algorithms for solving 
direct scattering problems related to acoustic wave scattering by 
soft and hard obstacles of arbitrary shapes. It gives an error 
estimate for solving the direct scattering problem. It suggests 
a numerical method for finding the shape of a starshaped obstacle from the 
scattering data.

\end{abstract}

%\end{titlepage}

\section{Introduction}
Consider a bounded domain $D \subset \R^n$, $n = 3$ with a boundary $S$. The
exterior domain is $D^\prime = \R^3 \backslash D$. Assume that $S$ is
smooth and starshaped, that is, its equation can be written as
$$r= f(\alpha), \eqno{(1.1)}$$
where $\alpha \in S^2$ is a unit vector
and $ S^2$ denotes the unit sphere in $R^3$. Smoothness of $S$ is used
in (4.6) below. For solving the direct scattering problem by the method 
described in 
the beginning of Section 2, the boundary $S$ can be Lipschitz.
The acoustic wave scattering problem by a soft obstacle $D$ 
consists in
finding the (unique) solution to the problem (1.2)-(1.3):
$$\left(\nabla^2 + k^2 \right) u=0 \hbox{\ in\ } D^\prime, \quad
  u = 0 \hbox{\ on\ } S, \eqno{(1.2)}$$
$$u=u_0 + A(\alpha^\prime, \alpha) \frac{e^{ikr}}{r} 
+ o
  \left(\frac{1}{r} \right), \quad r:=|x| \to \infty, \quad
  \alpha^\prime := \frac{x}{r}. \eqno{(1.3)}$$
Here $u_0:=e^{ik \alpha \cdot x}$ is the incident field,
$A(\alpha^\prime, \alpha)$ is called the scattering amplitude, its
k-dependence is not shown, $k>0$ is the wavenumber. Denote
$$A_\l (\alpha) := \int_{S^2} A(\alpha^\prime, \alpha)
  \overline{Y_\l (\alpha^\prime)} d\alpha^\prime, \eqno{(1.4)}$$
where $Y_\l (\alpha)$ are the orthonormal spherical harmonics,
$Y_\l = Y_{\l m}, -\l \leq m \leq \l$. Let $h_\l (r)$ be the spherical
Hankel functions, normalized so that
$h_\l (r) \sim \frac{e^{ikr}}{r}$ as $r \to +\infty$. Let the ball
$B_R := \{x : |x| \leq R\}$ contain $D$.

In the region $r> R$ the solution to (1.2) - (1.3) is:
$$u(x, \alpha) = e^{ik\alpha \cdot x} + 
\sum^\infty_{\l =0} A_\l (\alpha)\psi_\l, \quad
\psi_\l:= Y_\l (\alpha^\prime) h_\l (kr), \quad r > R,\quad  
\alpha^\prime =
  \frac{x}{r}, \eqno{(1.5)}$$
summation includes summation with respect to $m$, $-\l \leq m \leq \l$,
and $A_\l (\alpha)$ are defined in (1.4).

Rayleigh conjecture (RC):  the series (1.5) converges up to the 
boundary $S$ (originally RC dealt with periodic structures, gratings).
This conjecture is wrong \cite{1}, [3], [4]. For example, if $n=2$ 
and
$D$ is an ellipse, then the series analogous to (1.5) converges in the 
region
$r >a$, where $2a$ is the distance between the foci of the ellipse \cite{1}.
In the engineering literature there are numerical algorithms, based on the
Rayleigh conjecture. 
Our aim is to give a formulation of a modified Rayleigh conjecture (MRC)
which is correct and can be used in numerical solution of the direct and 
inverse scattering problems. We discuss the Dirichlet condition but 
similar argument is applicable to the Neumann boundary 
condition, corresponding to acoustically hard obstacles.

Fix $\epsilon >0$, an arbitrary small number.

{\bf Lemma 1.1.} {\it There exist $L=L(\epsilon)$ and 
$c_\l=c_\l(\epsilon)$
such that }
$$ ||u_0+\sum_{\l=0}^{L(\epsilon)}c_\l(\epsilon)\psi_l||_{L^2(S)} \leq 
\epsilon.
\eqno{(1.6)}$$ 

If (1.6) and the boundary condition (1.2) hold, then
$$ ||v_{\epsilon}-v||_{L^2(S)}\leq \epsilon,  \quad 
v_{\epsilon}:=\sum_{\l=0}^{L(\epsilon)}c_\l(\epsilon)\psi_l.
\eqno{(1.7)}$$

{\bf Lemma 1.2.} {\it If (1.7) holds then
$$ ||v_{\epsilon}-v||=O(\epsilon) \quad \epsilon \to 0, \quad
\eqno{(1.8)}$$
where $||\cdot||:= ||\cdot||_{H_{loc}^m(D')}+||\cdot||_{L^2(D'; 
(1+|x|)^{-\gamma})}$, $\gamma >1$, $m>0$ is an arbitrary integer,
and $H^m$ is the Sobolev space.} 

In particular, (1.8) implies
$$ ||v_{\epsilon}-v||_{L^2(S_R)}=O(\epsilon) \quad  \epsilon \to 0. 
\eqno{(1.9)}$$

{\bf Lemma 1.3.} {\it One has:}
$$ c_\l(\epsilon) \to A_\l(\alpha) \, \forall \l, \quad \epsilon \to 0.
\eqno{(1.10)}$$

The modified Rayleigh conjecture (MRC) is formulated as a theorem, which
follows from the above three lemmas:

{\bf Theorem 1 (MRC):} {\it For an arbitrary small $\epsilon>0$ there 
exist
$L(\epsilon)$ and $c_\l(\epsilon), 0\leq \l \leq L(\epsilon)$,
such that (1.6), (1.8) and (1.10) hold.}

The difference between RC and MRC is: (1.7) does not hold if one replaces 
$v_\epsilon$ by $\sum_{\l=0}^L A_\l(\alpha)\psi_\l$, and let $L\to 
\infty$ (instead of letting $\epsilon \to 0$).

For the Neumann boundary condition one minimizes
$ ||\frac {\partial [u_0+\sum_{\l=0}^{L}c_\l\psi_\l]}{\partial 
N}||_{L^2(S)}$
with respect to $c_\l$. Analogs of Lemmas 1.1-1.3 are valid and their 
proofs are essentially the same.

In Section 2 we discuss the usage of MRC in solving the direct scattering 
problem, in Section 3 its usage  in solving the inverse scattering
problem, and in Section 4 proofs are given.

\section{Direct scattering problem and MRC.} %2
The direct problem consists in finding the scattered field $v$
given $S$ and $u_0$. To solve it using MRC, fix a small $\epsilon >0$
and find $L(\epsilon)$ and $c_\l(\epsilon)$ such that 
(1.6) holds. This is possible by Lemma 1.1 and can be done numerically
by minimizing $||u_0+\sum_0^Lc_\l\psi_\l||_{L^2(S)}:=\phi 
(c_1,.....,c_L)$. If the minimum of $\phi$ is larger than $\epsilon$, 
then increase $L$ and repeat the minimization. Lemma 1.1 guarantees the
existence of such $L$ and $c_\l$ that the minimum is less than $\epsilon$.
Choose the smallest $L$ for which this happens and
define $v_\epsilon:=\sum^L_{\l = 0} c_\l \psi_\l (x)$. Then
$v_\epsilon$ is the approximate solution to the direct scattering problem
with the accuracy $O(\epsilon)$ in the norm $||\cdot||$ by Lemma 1.2. 

In [6] representations of $v$ and $v_\epsilon$ are
proposed, which greatly simplified minimization of $\phi$.
Namely, let $\Psi_\l$ solve problem 

$$\left(\nabla^2 + k^2 \right) \Psi_\l=0 \hbox{\ in\ } D^\prime, \quad
  \Psi_\l = f_\l \hbox{\ on\ } S,
\eqno{(2.1)}$$
and $\Psi_\l$ satisfies the radiation condition. Here $\{f_\l\}_{\l\geq 
0}$
is an arbitrary orthonormal basis of $L^2(S)$.
Denote
$$v(x) := \sum^\infty_{\l =0} c_\l \Psi_\l (x), \quad
  u(x) := u_0 + v(x), \quad c_l:=(-u_0, f_\l)_{L^2(S)}.
  \eqno{(2.2)}$$

The series (2.2) on $S$ is a Fourier series which converges
in $L^2(S)$.
It converges pointwise in $D'$ by the argument given in the proof of Lemma 
1.2. 
A possible choice of $f_\l$ for star-shaped $S$ is $f_\l= Y_\l/\sqrt{w}$ 
where 
$w:=dS/d\alpha$.
Here $dS$ and $d\alpha$ are respectively the elements of the surface areas 
of the surface $S$ and of the unit sphere $S^2$.

\section{Inverse scattering problem and MRC.}

 Inverse obstacle scattering problems IOSPa)
and  IOSPb) consist of finding $S$
and the boundary condition on $S$ from 
the knowledge of:
 
IOSPa): the scattering data $A(\alpha', \alpha, k_0)$ for all
$\alpha', \alpha \in S^2$, $k=k_0>0$ being fixed, 

or, 

IOSPb):  $A(\alpha', \alpha_0, k)$, known for all $\alpha' \in S^2$ 
and all $k>0$, $\alpha=\alpha_0 \in S^2$ being fixed.

Uniqueness of the solution to  IOSPa) is proved 
by A.G.Ramm (1985) for the Dirichlet, Neumann and Robin boundary 
conditions, and of  IOSPb) by M.Schiffer (1964), who  
assumed a priori the Dirichlet boundary condition. The proofs are given
in [4]. A.G.Ramm has also proved that not only $S$ but the boundary 
condition as well is uniquely defined by the above data in both cases,
and gave stability estimates for the solution to IOSP [9].
Later he gave a different method of proof of the uniqueness theorems
for these problems which covered the rough boundaries (Lipschitz and much
rougher boundaries: the ones with finite perimeter [8], see also [10].
In [11] the uniqueness theorem for the solution of inverse 
scattering problem is proved for a wide class of 
transmission problems. It is proved that not only the discontinuity 
surfaces of the refraction coefficient but also the coefficient itself
inside the body and the boundary conditions across these surfaces
are uniquely determined by the fixed-frequency scattering data.
For any strictly convex, smooth, reflecting obstacle $D$ analytical 
formulas
for finding $S$ from the high-frequency asymptotics of the scattering 
amplitude are proposed by A.G.Ramm, who gave error estimates of his 
inversion formula also [4].
The uniqueness theorems in the above references hold if the scattering 
data are given not for all $\alpha', \alpha \in S^2$, but only
for $\alpha'$ and $ \alpha$ in arbitrary small solid angles,
i.e., in arbitrary small open subsets of $S^2$.
The inverse scattering problem with the data $\alpha' \in S^2,$
$k=k_0$ and $\alpha=\alpha_0 $ being fixed, is open. If a priori
one knows that $D$ is sufficiently small, so that $k_0>0$
is not a Dirichlet eigenvalue of the laplacian in $D$, then uniqueness of 
the solution with the above non-overdetermined data holds
(by the usual argument [4]). There are many parameter-fitting 
schemes for solving IOSP, [13], see also [5].

Let us describe a new such scheme, based on MRC, its idea is similar to 
the one in [7]. 
Suppose that the scattered field $v$ is observed on a sphere $S_R$.
Calculate $c_\l:=(v,Y_\l)_{L^2(S^2)}/h_\l(kR)$. If $v$ is known exactly, 
then $c_\l=A_\l(\alpha)$. If $v_\delta$ are noisy data,
$||v-v_\delta||_{L^2(S_R)}\leq \delta$, then $c_\l=c_{\l \delta}$.
Choose some $L$, say $L=5$, and find $r=r(\alpha')$ as a positive root 
of the equation $u_0+v_L:=e^{ik\alpha \cdot \alpha' r}+
\sum_{\l=0}^L c_{\l \delta}\psi_\l(kr,\alpha'):=p(r,\alpha', \alpha, 
k)=0$.
Here $\alpha'$ and $k>0$ are fixed, and we are looking for the root
$r=r(\alpha')$ which is positive and stable under changes of $k$ and 
$\alpha$. In practice equation $p(r,\alpha', \alpha, k)=0$
may have no such root, the root may have small
imaginary part. If for the chosen $L$ 
such a root (that is, a root which is positive, or has a small
imaginary part, and stable with respect to changes of $k$ and $\alpha$) is 
not found, then increase $L$,
and/or decrease $L$, and repeat the search of the 
root. 
Stop the search at a smallest $L$ for 
which such a root is found.
The MRC justifies this method: for a suitable $L$ the function 
$p(r,\alpha', \alpha, k)$ is approximately equals zero on $S$, that is, 
for
$r=r(\alpha')$, and this $r(\alpha')$ does not depend on $k$ and $\alpha$.
Moreover, by the uniqueness theorem for IOSPa) and IOSPb)
there is only one such $r=r(\alpha')$. Numerically one expects to find 
a root of the equation $p(r,\alpha',k)=0$ which is close to positive
semiaxis $r>0$ and stable with respect to changes of $k$ and $\alpha$.

If one uses the above scheme for solving the inverse scattering problem
for an acoustically hard body (the Neumann boundary condition on $S$),
then one gets not a transcendental equation $p(r,\alpha', \alpha, k)=0$
for finding the equation of $S$, $r=r(\alpha')$, but a differential 
equation for $r=r(\alpha')$, which comes from the equation
$\frac {\partial p(r,\alpha', \alpha, k)}{\partial N}=0$ 
at $r=r(\alpha')$. One has to write the normal derivative on $S$
in spherical coordinates and then substitute  $r=r(\alpha')$
into the result to get a differential equation for the unknown function 
 $r=r(\alpha')$. For example, if $n=2$ (the two-dimensional case),
then the role of $\alpha'$ plays the polar angle $\varphi'$ and the 
equation for $r=r(\varphi')$ takes the form $\frac {dr}{d\varphi'}=
(r^2 \frac {dp}{dr}/\frac {dp}{d\varphi'})|_{r=r(\varphi')}$.

\section{Proofs.}

{\bf Proof of Lemma 1.1.}   This Lemma follows from the results in [4], 
(p.162, Lemma 1). 

\nd{\bf Proof of Lemma 1.2.} 
By Green's formula one has
$$ v_\epsilon (x) = -\int_S v_\epsilon (s) G_N (x,s) ds, \quad
  \| v_\epsilon (s)+u_0 \|_{L^2(S)} < \epsilon, 
\eqno{(4.1)}$$
where $G$ is the Dirichlet Green's function of the Laplacian in $D^\prime$:
$$\left(\nabla^2 + k^2 \right)G =-\delta (x-y) \hbox{\ in\ } D^\prime,
  \quad G=0 \hbox{\ on\ } S, \eqno{(4.2)}$$
$$\lim_{r \to \infty} \int_{|x| = r} \left|\frac{\partial G}{\partial |x|} 
-  ik G \right|^2 ds = 0 \eqno{(4.3)}$$
From (4.1) one gets (1.3) with $H^m_{loc}(D')-$norm immediately by the 
Cauchy inequality,
and with the weighted norm from the estimate
$$\left|G_N (x,s)\right| \leq \frac{c}{1 + |x|}, \quad |x| \geq R,
  \eqno{(4.4)}$$
and from local elliptic estimates for $w_\epsilon:=v_\epsilon -v$, which 
imply that
$$\|w_\epsilon \|_{L^2(B_R\backslash D)} \leq c\epsilon. 
\eqno{(4.5)}$$
Let us recall the elliptic estimate we use.
Let  $D'_{R}:=B_R\backslash D$,
$S_R$ be the boundary of $B_R$, and choose $R$ such that $k^2$ is not
a Dirichlet eigenvalue of $-\Delta$ in $D'_{R}$.
 The elliptic estimate we have used is (\cite{2'}, p.189):
$$\|w_\epsilon \|_{H^m(D'_{R})} \leq c [||(\Delta 
+k^2)w_\epsilon||_{H^{m-2}(D'_{R})}+
||w_\epsilon||_{H^{m-0.5}(S_R)} + ||w_\epsilon||_{H^{m-0.5}(S)}].
\eqno{(4.6)}$$ 
Take $m=0.5$ in (4.7), use the equation
$(\Delta+k^2)w_\epsilon=0$ in $D'$,  the estimate 
$||w_\epsilon||_{H^{m}(S_R)}=O(\epsilon)$, proved above,
the estimate $||w_\epsilon||_{H^{0}(S)}=O(\epsilon)$,
and get (1.8).
Lemma 1.2 is proved.
\qed

\nd{\bf Proof of Lemma 1.3.}
Lemma 1.2 yields convergence of $v_\epsilon$ to $v$ in the norm 
$||\cdot||$. In particular, $||v_\epsilon-v||_{L^2(S_R)}\to 0$
as $\epsilon \to 0$. On $S_R$ one has $v=\sum_{\l=0}^\infty 
A_\l(\alpha)\psi_\l$ and $v_\epsilon=\sum_{\l=0}^{L(\epsilon)}c_\l 
\psi_\l$. Multiply $v_\epsilon (R, \alpha')-v(R, \alpha')$ by
$\overline {Y_\l(\alpha')}$, integrate over $S^2$ and then let $\epsilon 
\to 0$. The result is (1.10).
\qed

\end{document}